\documentclass[9.5pt,letterpaper]{article}
\pdfoutput=1 

\usepackage[utf8]{inputenc} 
\usepackage{lipsum} 

\usepackage[top=2.75cm, bottom=2.5cm, left=3.5cm, right=3.5cm,
           heightrounded,marginparwidth=1.5cm, marginparsep=1cm]{geometry} 

\usepackage{changepage} 

\usepackage[shortlabels]{enumitem} 

\usepackage[square,numbers,merge,comma,sort&compress]{natbib} 
\makeatletter
\def\NAT@spacechar{\,}  
\makeatother

\usepackage{amsmath,amssymb,amsfonts,amsthm}
\usepackage{mathtools} 
\usepackage{old-arrows}
\usepackage[scale=0.7]{miama} 
\usepackage{etoolbox}
\usepackage{slashed,cancel}
\usepackage{comment}   
\usepackage{relsize}   
\usepackage{setspace}  
\usepackage{moresize}  
\usepackage{epsfig}
\usepackage{latexsym}
\usepackage{mathrsfs} 
\usepackage{calligra,aurical} 
\usepackage{calc}
\usepackage{float}
\usepackage{appendix}
\usepackage{xargs}
\usepackage{extarrows}
\usepackage{empheq}
\usepackage{url}
\usepackage{hang} 

\usepackage[blocks]{authblk}  

\usepackage[fulladjust]{marginnote}%

\usepackage{graphicx} 
\graphicspath{{Figures/}} 

\usepackage[labelsep=colon]{caption}  
\usepackage[labelsep=colon,aboveskip=10pt,belowskip=10pt]{subcaption}
\usepackage{array,multirow,makecell,booktabs}  
\newcolumntype{X}[2]{>{\centering\arraybackslash$}#1{#2\linewidth}<{$}}
\newcolumntype{R}[1]{>{\raggedleft\arraybackslash$}m{#1\linewidth}<{$}}
\newcolumntype{L}[1]{>{\raggedright\arraybackslash}m{#1\linewidth}}

\makeatletter
\renewcommand\mcell@classz{\@classx
   \@tempcnta \count@
   \prepnext@tok
   \@addtopreamble{
      \ifcase\@chnum
         \hfil
         \mcell@agape{\d@llarbegin\insert@column\d@llarend}\hfil \or
         \hskip1sp
         \mcell@agape{\d@llarbegin\insert@column\d@llarend}\hfil \or
         \hfil\hskip1sp
         \mcell@agape{\d@llarbegin \insert@column\d@llarend}\or
         \mcell@agape{$\vcenter
         \@startpbox{\@nextchar}\insert@column\@endpbox$}\or
         \mcell@agape{\vtop
         \@startpbox{\@nextchar}\insert@column\@endpbox}\or
         \mcell@agape{\vbox
         \@startpbox{\@nextchar}\insert@column\@endpbox}%
      \fi
      \global\let\mcell@left\relax\global\let\mcell@right\relax
    }\prepnext@tok}
\makeatother

\usepackage{titlesec}

\titleformat{\section}{\normalfont\large\bfseries}{\thesection}{1em}{}
\titleformat{\subsection}{\normalfont\normalsize\bfseries}{\thesubsection}{0.75em}{}
\titleformat{\subsubsection}{\normalfont\normalsize\bfseries}{\thesubsubsection}{0.75em}{}

\titlespacing*{\section}{0pt}%
                {4ex plus 1ex minus .5ex}{1.75ex plus .25ex minus .25ex}
\titlespacing*{\subsection}{0pt}%
                {3.5ex plus 1ex minus .5ex}{1.25ex plus .2ex minus .2ex}
\titlespacing*{\subsubsection}{0pt}%
                {2.5ex plus 0.75ex minus .2ex}{0.75ex plus .15ex minus .15ex}
\titlespacing*{\paragraph}{0pt}%
                {1.85ex plus 0.5ex minus .15ex}{1em}

\usepackage{titletoc}

\addtocontents{toc}{\addvspace{-0.75em}}  

\titlecontents{section}
  [1.25em] {\addvspace{0.7em plus 0pt}\small}
  {\thecontentslabel\hspace{0.75em}}{}
  {\hspace{0.5em}\titlerule*[0.5em]{.}\contentspage}
  [\addvspace{0.0em plus 0pt}]

\titlecontents{subsection}
  [2.75em] {\addvspace{0.075em plus0pt}\fns}
  {\thecontentslabel\hspace{0.75em}}{\thecontentslabel\hspace{0.75em}}
  {\hspace{0.5em}\titlerule*[0.5em]{.}\small\contentspage}
  [\addvspace{0.075em plus 0pt}]

\usepackage{environ}
\makeatletter
\NewEnviron{subalign}[1]{
\begin{subequations}\label{#1}
\begin{align} \BODY \end{align}
\end{subequations}      }
\makeatother
\makeatletter
{\begingroup%
\setlength{\abovedisplayskip}{10pt plus 4pt minus 9pt}%
\setlength{\abovedisplayshortskip}{0pt plus 2pt minus 2pt}%
\setlength{\belowdisplayskip}{12pt plus 3pt minus 9pt}%
\setlength{\belowdisplayshortskip}{7pt plus 3pt minus 4pt}%
\begin{subequations}%
}%
{\end{subequations}\ignorespacesafterend%
\endgroup}%
\makeatother

\makeatletter
{\begingroup%
\setlength{\abovedisplayskip}{{#1}pt plus 2pt minus 9pt}%
\setlength{\abovedisplayshortskip}{0pt plus 0pt minus 2pt}%
\setlength{\belowdisplayskip}{{#2}pt plus 3pt minus 9pt}%
\setlength{\belowdisplayshortskip}{7pt plus 3pt minus 4pt}%
\begin{subequations}%
}%
{\end{subequations}\ignorespacesafterend%
\endgroup}%
\makeatother

\makeatletter
{\begingroup%
\setlength{\abovedisplayskip}{{#1}pt plus 3pt minus 9pt}%
\setlength{\abovedisplayshortskip}{0pt plus 3pt}%
\setlength{\belowdisplayskip}{{#2}pt plus 3pt minus 9pt}%
\setlength{\belowdisplayshortskip}{7pt plus 3pt minus 4pt}%
\begin{equation}%
}%
{\end{equation}\ignorespacesafterend%
\endgroup}%
\makeatother

\makeatletter
{%
\begin{equation}%
\begin{split}{#1}%
}%
{\end{split}%
\end{equation}\ignorespacesafterend%
}%
\makeatother

\usepackage[svgnames]{xcolor}
\definecolor{Green}{rgb}{0.05, 0.45, 0.25}
\definecolor{dogwoodrose}{rgb}{0.8, 0.1, 0.55}
\definecolor{RRed}{rgb}{0.7, 0.1, 0.525}

\usepackage{bm}  
\usepackage{dsfont}  

\DeclareMathAlphabet{\mathpzc}{OT1}{pzc}{m}{it}
\DeclareMathAlphabet{\mathcal}{OMS}{cmsy}{m}{n}
\DeclareSymbolFontAlphabet{\Scr}{rsfs}
\DeclareMathAlphabet{\mathbold}{U}{BOONDOX-ds}{m}{n}
\SetMathAlphabet{\mathbold}{bold}{U}{BOONDOX-ds}{b}{n}
\DeclareMathAlphabet{\mathcalboondox}{U}{BOONDOX-calo}{m}{n}
\SetMathAlphabet{\mathcalboondox}{bold}{U}{BOONDOX-calo}{b}{n}
\DeclareMathAlphabet{\mathbcalboondox}{U}{BOONDOX-calo}{b}{n}

\newcommand\eqlinkcol{RRed}



\usepackage[breaklinks=true,backref=page]{hyperref}
\hypersetup{
    bookmarks=true,         
    pdfmenubar=true,        
    pdffitwindow=false,     
    pdfpagemode={UseNone},
    pdfstartview={FitH},
    pdfauthor={},     
    pdfsubject={},   
    pdfcreator={},   
    pdfproducer={},  
    colorlinks=true,
    bookmarks=true,
    bookmarksnumbered=true,
    plainpages,
    a4paper,
    linktoc=page,
    citecolor=blue,
    filecolor=black,
    linkcolor=\eqlinkcol,
    urlcolor=Green,
}
\renewcommand*{\backref}[1]{}
\renewcommand*{\backrefalt}[4]{%
\ifcase #1 %
\relax
\or
~{\small [\textsc{p.~\fns{\!#2}}]}
\else
~{\small [\textsc{p.~\fns{\!#2}}]}%
\fi}

\usepackage{footnotebackref}
\usepackage{hypernat} 
\usepackage[noabbrev,capitalise]{cleveref}

\def\+{~+~}
\def\-{~-~}
\def\={\:=\:}
\def\'{``}
\def\*{{}^*}
\newcommand\fns{\footnotesize}

\newcommand\qqquad{\quad\quad\quad}

\newcommand{\ms}{\mathsmaller}

\newcommand{\ml}{\mathlarger}

\providecommand{\abs}[1]{\left\lvert#1\right\rvert}

\DeclareMathOperator\arctanh{arctanh}

\newcommand\veps{\varepsilon}
\newcommand\w{\omega}
\newcommand\Id{\mathds{1}}

\newcommand\vF{v_\textsc{f}}
\newcommand\ustar{u_\ms{\star}}
\newcommand\C{\mathcal{C}}
\newcommand\B{\mathscr{B}}
\newcommand\Iplus{I^{{}^\ms{(+)}}}
\newcommand\Iminus{I^{{}^\ms{(-)}}}
\newcommand\I{\mathcal{I}}
\newcommand\V{\mathpzc{V}}
\newcommand\PP{\mathcal{P}}
\newcommand\X{\mathpzc{X}}
\newcommand\LDOS{\rho_{{}_\textsc{s}}}
\newcommand\EF{E_{\textsc{f}}}

\newcommandx{\tcR}[1]{\textcolor{Crimson}{#1}}
\newcommandx{\tts}[1]{\text{\textsmaller{#1}}}
\newcommandx{\dd}[1][1=\mu,usedefault]{\partial_{#1}}
\newcommandx{\ddu}[1][1=\mu,usedefault]{\partial^{#1}}
\newcommandx{\subm}[2][1=p,2=A,usedefault]{{#1}_{\!\mathsmaller{#2}}}
\newcommandx{\subt}[2][1=p,2=A,usedefault]{{#1}_\text{\textsmaller{#2}}}
\newcommandx{\supm}[2][1=p,2=A,usedefault]{{#1}^{\!\mathsmaller{#2}}}
\newcommandx{\supt}[2][1=p,2=A,usedefault]{{#1}^\text{\textsmaller{#2}}}
\newcommandx{\subpt}[3][1=p,2=A,3=B,usedefault]{{#1}^\text{\textsmaller{#3}}_\text{\textsmaller{#2}}}
\newcommandx{\subpm}[3][1=p,2=A,3=B,usedefault]{{#1}^{\mathsmaller{#3}}_{\mathsmaller{#2}}}
\newcommandx{\sh}[1][1=\alpha,usedefault]{\sinh\left(#1\right)}
\newcommandx{\ch}[1][1=\alpha,usedefault]{\cosh\left(#1\right)}
\newcommandx{\sech}[1][1=\alpha,usedefault]{\mathrm{sech}\left(#1\right)}
\newcommandx{\cosech}[1][1=\alpha,usedefault]{\mathrm{cosech}\left(#1\right)} \newcommandx{\LCTd}[4][1=\mu,2=\nu,3=\rho,4=\sigma,usedefault]{\veps_{#1#2#3#4}}
\newcommandx{\LCTu}[4][1=\mu,2=\nu,3=\rho,4=\sigma,usedefault]{\veps^{#1#2#3#4}}
\newcommandx{\gmat}[2][1=\mu,2=\nu,usedefault]{g_{{#1}{#2}}}
\newcommandx{\gmatinv}[2][1=\mu,2=\nu,usedefault]{g^{{#1}{#2}}}
\newcommandx{\spc}[3][1=\mu,2=a,3=b,usedefault]{{\w_{#1}}^{\!\!{#2}{#3}}}
\newcommandx{\Conn}[3][1=\mu,2=\nu,3=\lambda,usedefault]{{\Gamma_{{#1}{#2}}}^{\!\!#3}}
\newcommandx{\viel}[2][1=\mu,2=a,usedefault]{{e_{#1}}^{\!#2}}
\newcommandx{\vielinv}[2][1=a,2=\mu,usedefault]{{e_{#1}}^{#2}}
\newcommandx{\vieluu}[2][1=\mu,2=a,usedefault]{e^{#1#2}}
\newcommandx{\Rdduu}[4][1=\mu,2=\nu,3=a,4=b,usedefault]{{R_{{#1}{#2}}}^{{#3}{#4}}}
\newcommandx{\DD}[1][1=\mu,usedefault]{\mathcal{D}_{#1}}
\newcommandx{\gammafl}[1][1=\gamma,usedefault]{\mkern2mu\check{\mkern-2mu#1}}

\newcommandx{\overbar}[1]{\mkern
1.5mu\overline{\mkern-2.0mu#1\mkern-2.0mu}\mkern 1.5mu}
\newcommandx{\overbarcal}[1]{\mkern                   6.0mu\overline{\mkern-5.5mu#1\mkern-1.0mu}\mkern 1.5mu}  

\makeatletter
\normalsize
\makeatother

\DeclareFixedFont\trfont{OT1}{phv}{b}{sc}{11}

\title{%
       \vspace{-1.0cm}
       \centering\boldmath\LARGE\bfseries%
       Negative-curvature spacetime solutions for graphene
       \bigskip
       }

\medskip

\author{\small\textsc{Antonio Gallerati}%
\vspace{0.15em}%
}
\affil{%
\makebox[\textwidth][c]{Politecnico di Torino, Dipartimento di Scienza Applicata e Tecnologia, corso Duca degli Abruzzi 24, 10129 Torino, Italy}
}
\affil{Istituto Nazionale di Fisica Nucleare, Sezione di Torino, via Pietro Giuria 1, 10125 Torino, Italy%
}
\affil{\href{mailto:antonio.gallerati@polito.it}{\texttt{antonio.gallerati@polito.it}}%
}

\date{}

\makeatletter
\patchcmd{\@maketitle}{\begin{center}}{\begin{adjustwidth}{-0.25in}{-0.25in}\begin{center}}{}{}
\patchcmd{\@maketitle}{\end{center}}{\end{center}\end{adjustwidth}}{}{}
\makeatother

\begin{document}

\maketitle


\begin{abstract}
{\noindent%
We provide a detailed analysis of the electronic properties of graphene-like materials with charge carriers living on a curved substrate, focusing in particular on constant negative-curvature spacetime. An explicit parametrization is also worked out in the remarkable case of Beltrami geometry, with an analytic solution for the pseudoparticles modes living on the curved bidimensional surface. We will then exploit the correspondent massless Dirac description, to determine how it affects the sample local density of states.
}%
\end{abstract}


\bigskip

\tableofcontents

\pagebreak


\section{Introduction}\label{sec:Intro}
The recent developments in material science provide a new connection between condensed matter and quantum electrodynamics models. In particular, the study of the physics of carbon-based materials like graphene opens a window on the possibility of a direct observation of quantum behaviour in the curved background of a solid state system \cite{katsnelson2007graphene,geim2007rise,CastroNeto:2009zz}.
A graphene sheet is a bidimensional system of carbon atoms arranged in a honeycomb lattice of one-atom thickness, one of the closest possible real two-dimensional objects.
In 1984 Semenoff formulated the hypothesis that graphene could realize the physics of two dimensional massless Dirac fermions \cite{Semenoff:1984dq}, this property discriminating graphene from other 2D system. Graphene crystals were then produced in 2004 as single carbon atom layers \cite{novoselov2004electric,novoselov2005twodimato}.\par
As we will discuss in detail, graphene and other 2D materials realize the physics of spinorial fields, whose Dirac properties emerge due to the structure of the space (lattice) with which the charge carriers interact.
The peculiar sheet structure then determines a natural description of its electronic properties in terms of massless pseudoparticles, giving the possibility to study quasi-relativistic particle behaviour at sub-light speed regime%
\footnote{%
our framework turns out to be the analog of a relativistic system, with
characteristic limiting velocity given by the Fermi velocity  $\vF$ rather than the speed of light $c$ \,(for graphene $\vF\sim\tfrac{c}{300}$)
}
\cite{Novoselov:2005kj,Gusynin:2006ym,CastroNeto:2009zz}.
A natural suggestion turns out to be that the geometric curvature of the two-dimensional sample, combined with the mentioned special relativistic-like behaviour, naturally leads to a general relativistic-like description for our pseudoparticles, which will then be regarded as Dirac fields in a 1+2 dimensional curved spacetime background \cite{Birrell:1982ix,Brill:1966tia,Wald:1984rg,Cortijo:2006mh,Cortijo:2006xs,Vozmediano:2010zz}.
This gives us a real framework to study what is believed to be (as close as possible) a quantum field in a curved spacetime, with measurable effects pertaining to the electronic structure of the sample itself \cite{Cortijo:2006mh,Cortijo:2006xs,Gorbar:2007kd,Boada:2010sh,Gallerati:2018dgm}, so that the understanding of 2D Dirac materials properties is important in condensed matter as well as in theoretical high energy physics \cite{geim2007rise,katsnelson2007graphene,Gallerati:2018dgm}.\par
The massless formulation is in general robust, since it emerges at the level of non-interacting system, the vanishing quasiparticle mass (gapless spectrum) protected by the combination of parity and time-reversal symmetries. In general, interactions are not very efficient in introducing a gap and/or modifying the quasiparticle behavior \cite{CastroNeto:2009zz,Kotov:2010yh}.\par
In the context of high energy physics, the emergence of intrinsic and extrinsic curvature in graphene-like materials can be used to investigate the fundamental physics of the quantum Dirac dynamics in curved spacetimes, as well as to probe certain quantum gravity scenarios \cite{novello2002artificial,Barcelo:2005fc}.
This formulation follows a \emph{bottom-up} approach, where suitable condensed matter systems provide analogues of gravitational effects so that the propagation of quantum fields is dictated by an effective metric, taking then advantage of mathematical tools from Einstein gravity (or extensions of the latter). The underlying idea is that suitable variants of this analogue models can be used as frameworks for different analysis and formulations of quantum gravity theories, gaining new insights into the corresponding problems.\par
There are also some theoretical results that conjecture the use of graphene to have alternative (unconventional) realizations of Supersymmetry \cite{Alvarez:2011gd,Alvarez:2013tga}, the latter being instrumental in describing the properties of graphene-like materials at the Dirac points, exploiting an holographic \emph{top-down} approach, the substrate description coming from a well-defined geometric formulation of a suitable gravitational model \cite{Andrianopoli:2019sip,Gallerati2020supersymmetric}.

\paragraph{Continuum limit and spacetime geometry.}
The detailed study of suitable curved configurations can highlight the peculiar properties of the charge carriers, derived from the discussed massless Dirac description in a curved background \cite{osipov2005electronic,kolesnikov2006continuum,morpurgo2006intervalley,lee2009surface}: the choice of the geometry, the corresponding parametrization and the quantization of some physical quantities, can lead to characteristic observable effects.\par
In general, we can state that Dirac physics can be realized for our quasiparticles considering low-lying energy excitations: if we consider energy ranges below $E_\ell\,\sim\,\vF/\ell\,\sim\,4.6\,\mathrm{eV}$, the electrons wavelength is large compared to the lattice spacing $\ell\sim0.142\,\mathrm{nm}$, so that these charge carriers see the graphene sheet as a continuum, justifying the quantum description in 1+2 spacetime. Moreover, quasiparticles with large wavelength are sensitive to sheet curvature effects, claiming for a quantum field formulation in curved spacetime. In particular, this means that, in the continuum field approximation, we have to demand the charge carrier wavelengths to be bigger than the lattice typical dimension, $\lambda\,>\,2\pi\vF/E_\ell\,\sim\,2\pi\ell$\,.\par\smallskip
%
With the above prescriptions in mind, the challenge is now to find a suitable curved spacetime where it can be easier to probe and study the relativistic-like quasiparticles quantum behaviour. As we will see in Section \ref{sec:Negcurvspace}, the Beltrami pseudosphere \cite{Beltrami2017saggio} is a promising candidate where Dirac's equation in curved space can be solved analytically, providing an explicit expression for the Dirac spectrum and its effects on the electronic local density of states (LDOS).\par
From an historical point of view, the Beltrami surface has been conjectured to provide a promising spacetime framework where to observe an Hawking–Unruh effect \cite{Iorio:2011yz,Chen:2012uc}, one of the most interesting predicted phenomena of a quantum field theory in curved background \cite{Hawking:1974rv,Unruh:1976db}. The possible formation of a Rindler-type horizon in a Beltrami geometry could then lead to a characteristic thermal behaviour, related to the specific nature of quantum vacua and relativistic process of measurement \cite{Iorio:2012xg,Iorio:2013ifa,morresi2020exploring}. In Section \ref{sec:Analyticsol} we will provide an analytic expression for the Dirac modes of the charge carriers living in a Beltrami spacetime, so that experimental predictions related to the electronic structure of the corresponding graphene-like sample can be explicitly worked out.\par
Finally, we point out that, although we have primarily graphene in mind, many of the following considerations can be extended to other two-dimensional Dirac materials, including silicene, germanene, graphynes, several boron and carbon sheets, transition-metal oxides (TiO$_2$/VO$_2$), organic and organometallic crystals (MoS$_2$), artificial lattices (electron gases and ultracold atoms) \cite{Wang2015rare,Cahangirov2009twoandone,Malko2012competition,Xu2014twodim,Zhou2014semimetallic,Pardo2009half,Katayama2006pressure,Li2014gapless,Zhu2007simulation}.

\section{Dirac formalism}\label{sec:Dirac}
The quantum Dirac formulation introduced above emerges from the graphene lattice structure, where a unit cell is made of two adjacent atoms belonging to the two inequivalent, interpenetrating triangular sublattices. This means that we have two inequivalent sites per unit cell, the distinction related to their topological inequivalence. The single-electron wave function can be then conveniently arranged in a two-component Dirac spinor, so that the description of its electronic properties can be given in terms of massless Dirac pseudoparticles \cite{CastroNeto:2009zz}, the characterization being resistent to changes of the lattice preserving the topological structure.\par
In the reciprocal lattice space, the first Brillouin zone (FBZ) results in a structure with the same hexagonal form of the honeycomb lattice, rotated by a $\pi/2$ angle. The relativistic behavior of the charge carriers can be inferred, in the momentum space, from the linear dispersion relation between energy and quasimomentum at the corners of the FBZ. The latter can be divided into two topological inequivalent classes, since only two of the six vertices can be chosen to be independent, the remaining four connected to them by a reciprocal lattice vector; this means we can consider only two inequivalent corners, labeled \textbf{K}, \textbf{K'} (\emph{Dirac points}).
%

\paragraph{Substrate deformations and energy scales.}
If we consider a graphene layer with hexagonal lattice, every carbon atom has four electrons available for covalent bonds. Three of them form the $\sigma$-bonds with three different nearest neighbors (merging of atomic 2$s$ orbitals); these bonds define the elastic properties of the sheet. The fourth electron forms a covalent $\pi$-bond with one of the three neighbors (merging of atomic 2$p$ orbitals): being the latter $\pi$-bond much weaker than the former $\sigma$, the involved $\pi$-electrons become charge carriers that are much more free to hop, determining then the electronic properties of the sample.\par
If we want to construct the action that captures the physics of the $\pi$-electrons in the curved sheet, we need to study the possible deformations that can be encoded in the Dirac description. In the large wavelength regime \cite{Peres:2010mx,Cortijo:2006mh,Cortijo:2006xs}, we can find three kinds of deformation at work: extrinsic curvature, intrinsic curvature and strain \cite{CastroNeto:2009zz}.
The first deformation is an elastic effect that can be expressed, at first order, using derivatives of the strain tensor \cite{CastroNeto:2009zz,deJuan:2012hxm}; the second is an inelastic effect coming from the formation of disclination-type defects \cite{Kleinert1989gauge,Katanaev:1992kh}; the third is again an elastic deformation that takes into account effects that are proportional to the strain tensor (not to its derivatives), that turn out to work as potentials for a pseudo-magnetic field $B_\mu$ and a scalar potential $\Phi$ \cite{Guinea2008midgap,CastroNeto:2009zz,Levy2010strain,deJuan:2012hxm,Morozov2006strong,Stegmann:2015mjp}. \par
Since we are primarily interested in investigating the effects of curvature on the substrate quantum (electronic) properties, we shall focus on inelastic deformations, since elastic deformations cannot induce intrinsic sheet curvature \cite{CastroNeto:2009zz}. This means that we will have to introduce also the energy scale $E_{\!R}\,\sim\,\vF/R\,<\,E_\ell$\,, \,with $R>\ell$ \,and where $1/R^2$ is a measure of the intrinsic curvature. In fact, we want the curvature to be small if compared to the limiting value $1/\ell^2$\,: this, in turn, means that we can formulate our theory using a smooth metric, ruling also out the difficult bending of the strong $\sigma$-bonds. The previously introduced $E_\ell$ energy now corresponds to the high-energy regime for our formulation and, when we are within the $E_{\!R}$ energy range, the charge carriers are still sensitive to the global effects of curvature.\par
The above considerations suggest we should focus on electrons with longer wavelengths than the corresponding ones for the simple continuum approximation, ${\lambda\,>\lambda_R\,>\,\lambda_\ell}$ \;(with\, ${\lambda_\ell\,\sim\,2\pi\ell}$, \,${\lambda_R\,\sim\,2\pi R}$), so that our energies range is valid up to $E_{\!R}$: in this situation, the elastic properties of the sample (involving much larger energies, of the order of tens of eV) are decoupled from the $\pi$-electrons dynamics, governed by inelastic effects, and in our mathematical formulation we will neglect the contributions from $B_\mu$ and $\Phi$ \cite{Vozmediano:2008zz,deJuan2007charge}.

\paragraph{Topological defects.}
The formation of topological defects in 2D materials is the natural way in which the sample layer heals vacancies and other analogous lattice damages. Among those, disclinations, dislocations and Stone–Wales defects (special dislocation dipoles) were found to have the least formation energy and activation barrier, so that they result energetically favourable phenomenons \cite{Carpio2008dislocations,deJuan2010dislocations}.
If we consider graphene-like materials, disclinations and dislocations are the most important topological sample defects%
\footnote{%
we do not consider here impurities, Coulomb and resonant scattering or other issues mixing the Fermi points, with a corresponding chiral term in the action: we can assume that charge carriers mobility is not affected by the mentioned effects at these energy scales \cite{Peres:2010mx}. The local lattice aspects can be then disregarded and the inelastic effects will dominate, so that only intrinsic curvature must be taken into account (contributions from $B_\mu$ and $\Phi$ can be neglected) \cite{deJuan2007charge,Vozmediano:2008zz}
}, %
related, in the continuum limit, to curvature and torsion, respectively \cite{Kleinert1989gauge,seung1988defects}.\par
A disclination is a crystallographic defect associated with the violation of the (discrete) rotational symmetry. Positive (negative) disclinations are topological defects obtained by removing (adding) a semi-infinite wedge of material to an otherwise perfect lattice. If we consider a bidimensional hexagonal lattice, a disclination consists in the substitution of an hexagon with other polygons: it therefore manifests itself by the presence of an $n$-sided polygon, with $n\neq6$. If \,$3\leq n<6$, the associated singularity carries a positive intrinsic curvature, while, in the $n>6$ case, it carries a negative intrinsic curvature .
A dislocation defect appears in graphene-like lattices as a disclination pair (dipole of disclinations, usually pentagon–heptagon pairs); its effects manifest themselves in forms similar to those arising from curvature or elastic strain.\par
Inelastic phenomenons due to geometric curvature can be then generated by the presence of suitable sample topological defects \cite{Cortijo:2006mh,Cortijo:2006xs,Gupta:2008mn,Kleinert1989gauge,seung1988defects,nelson2002defects}.
In the following sections, we will deal in detail with the case of negative-curvature surfaces; the latter usually carry one of the described topological defects (for example, heptagon-disclination defects in hexagonal lattice). However, we will not have to face in detail the local microscopical substrate deformations since, as we already stated, we are working in continuum limit.

\subsection{Dirac equation}
Graphene-like flat substrates can be considered as 2D analogs of pseudorelativistic systems with characteristic velocity $\vF$\,. The dynamics of the charge carriers $\psi$ in the 1+2 dimensional flat spacetime can be then described, in the long wavelength continuum limit, by a massless Dirac action of the form
\begin{equation}
\mathcal{S}_0\=i\,\hbar\,\vF \int{d^{3}x\;\bar{\psi}\;\gammafl^a\,\dd[a]\psi}\;,\qqquad
\label{eq:flatDiracaction}
\end{equation}
where $a=1,2,3$ is the flat spacetime index and where, for the sake of notational simplicity, we have omitted spinorial indices. A set of three-dimensional (flat) $\gammafl$-matrices can be written in terms of the Pauli matrices as
\begin{equation}
\gammafl_a\=\big(\,i\,\sigma_3\,,\;\sigma_1\,,\;\sigma_2\,\big)\;.
\label{eq:gammaflat}
\end{equation}
One can easily verify that the $\gammafl^a=\eta^{ab}\,\gammafl_b$ matrices satisfy the standard Clifford algebra $\left\{\gammafl^a,\gammafl^b\right\}=2\,\eta^{ab}\,\Id$\,, where $\eta^{ab}$ is the inverse of the flat 1+2 dimensional Minkowski metric in the mostly plus convention, $\eta_{ab}=\mathrm{diag}(-1,1,1)$\,.

\paragraph{Curved space.}
Since we now want to include non-trivial intrinsic curvature effects, we are naturally led to the customary generalization in a curved spacetime of the action for massless Dirac spinors in 1+2 dimensions \cite{Wald:1984rg,Birrell:1982ix}
\begin{equation}
\mathcal{S}\=i\,\hbar\,\vF \int{d^{3}x\,\sqrt{g}\;\bar{\psi}\,\gamma^\mu\,\DD[\mu]\psi}\;,\qqquad
\label{eq:curvedDiracaction}
\end{equation}
where $\mu=0,1,2$ is now an index referring to the new curved spacetime with metric $\gmat$, and the factor $\sqrt{g}\equiv\sqrt{-\det(\gmat)}$ comes from the request of a diffeomorphic-covariant form of the action in the presence of curvature.
The curved $\gamma_\mu$ matrices are obtained from the constant $\gammafl_a$ matrices of the flat frame by the action of the vielbein $\viel$ (see App.\ \ref{app:Conventions}):
\begingroup%
\setlength{\abovedisplayshortskip}{5pt plus 3pt}%
\setlength{\abovedisplayskip}{5pt plus 3pt minus 4pt}
\setlength{\belowdisplayshortskip}{5pt plus 1pt}%
\setlength{\belowdisplayskip}{7pt plus 1pt minus 1pt}%
\begin{equation}
\gamma_\mu\=\viel\,\gammafl_a\;,
\end{equation}
\endgroup
while the inverse vielbein $\vielinv$ performs the transformation in the other direction. The gamma matrices with upper indices \,$\gamma^\mu=\gmatinv\,\gamma_\nu$\, satisfy the correspondent Clifford algebra relation in curved background, \,$\{\gamma^\mu,\,\gamma^\nu\}\=2\,\gmatinv\,\Id$\,.\par
The diffeomorphic covariant derivative is written as
\begingroup%
\setlength{\abovedisplayshortskip}{6pt plus 3pt}%
\setlength{\abovedisplayskip}{7pt plus 3pt minus 4pt}
\setlength{\belowdisplayshortskip}{5pt plus 1pt}%
\setlength{\belowdisplayskip}{6pt plus 1pt minus 1pt}%
\begin{equation}
\DD\=\dd+\Omega_\mu\=\dd+\frac{1}{4}\,\spc\,M_{ab}\;,
\label{eq:covder}
\end{equation}
\endgroup
where $M_{ab}=\frac12\,[\gammafl_a,\gammafl_b]$ are the Lorentz generators and $\spc$ defines the \emph{spin connection}, that can be seen as the gauge field of the local Lorentz group. Since we are working in a torsionless framework, $\spc$ and $\viel$ are not independent \cite{Eguchi:1980jx,Green:1987sp} and the former can be expressed in terms of the latter as
\begin{equation}
\spc \= \viel[\nu][a]\,\dd\vieluu[\nu][b]+\viel[\nu][a]\,\Conn[\mu][\lambda][\nu]\,\vieluu[\lambda][b]\;,
\end{equation}
where $\Conn$ is the affine connection
\begin{equation}
\Conn= \frac12\,\gmatinv[\sigma][\lambda]
       \left(\dd\gmat[\nu][\sigma]+\dd[\nu]\gmat[\mu][\sigma]-\dd[\sigma]\gmat\right)\;.
\end{equation}
If we consider the above \eqref{eq:covder}, $\Omega_\mu$ acts as a gauge field able to take into account all deformations of the geometric kind.\par
%
The equations of motion for the pseudorelativistic Dirac spinors coming from action \eqref{eq:curvedDiracaction} read%
\footnote{%
from now on we work in (pseudo)natural units: $\hbar=\vF=1$
}
\cite{Birrell:1982ix,Brill:1966tia,Cortijo:2006mh,Cortijo:2006xs,Vozmediano:2010zz}
\begin{equation}
i\,\gamma^\mu\,\DD\psi\=0\;,
\label{eq:Direqcurvmassless}
\end{equation}
that is, the generalized form of the massless Dirac equation in flat Minkowski spacetime obtained through the substitutions
\begingroup%
\setlength{\abovedisplayshortskip}{6pt plus 3pt}%
\setlength{\abovedisplayskip}{6pt plus 3pt minus 4pt}
\setlength{\belowdisplayshortskip}{5pt plus 1pt}%
\setlength{\belowdisplayskip}{6pt plus 1pt minus 1pt}%
\begin{equation}
\gammafl^a\:\rightarrow\:\gamma^\mu\,,
\qquad\;
\dd[a]\:\rightarrow\:\DD\;.\quad
\end{equation}
\endgroup
Summarizing, we have constructed the large wavelength continuum description
of the Dirac quasiparticles living on a graphene-like sheet, modelling curvature effects through the coupling of the Dirac fields to a curved spatial metric, thus obtaining the physical description of the charge carriers dynamics \cite{Cortijo:2006mh,Cortijo:2006xs,Vozmediano:2010zz,deJuan2007charge,Boada:2010sh,Gallerati:2018dgm}.
If we consider a topologically trivial, purely strained configuration (elastic membrane deformations), there are no relevant physical effects coming from the spin-connection. If instead the sample features a non-trivial intrinsic curvature, the spin connection dictates most of the physics for the Dirac fields. In particular, in our large wavelength continuum limit for the charge carriers, the spin connection can be associated with disclination-type defects inducing curvature, encoding the physics imposed by the geometric sheet deformation.\par
In the following sections we will consider an explicit parametrization of the curved membrane, working out an analytic solution of the Dirac equation in the corresponding curved background.

\section{Constant negative-curvature spaces}\label{sec:Negcurvspace}
Among the class of negative-curvature surfaces, an important role is played by the subset of surfaces having constant negative Gaussian curvature. When embedded into $\mathbb{R}^3$, those surfaces feature essential singularities \cite{Hilbert1933flachen}: as a consequence of this, we find that it is impossible to represent the whole Lobachevskian geometry on a real bidimensional surface, so that we are forced to restrict to mapping only a suitable stripe of the hyperbolic space \cite{Beltrami2017saggio}.\par
Another relevant point will concern explicit parametrization, so that hyperbolic (abstract) geometry can be expressed, together with the above singular boundaries, in terms of well defined coordinates. In this regard, we notice that the line element of any surface of constant negative Gaussian curvature can be reduced to the one of the Beltrami or the hyperbolic or the elliptic pseudospheres \cite{eisenhart1909treatise}. The advantages of the Beltrami surface are that an embedding parametrization can be given in terms of smooth, well-behaving single-valued functions and that it has only one well-defined singular boundary, corresponding to the maximal circle.\par
Below, we provide an embedding for the Beltrami pseudosphere in three spatial dimensions, the explicit parametrization given in terms of surface coordinates.

\subsection{Beltrami spacetime}\label{subsec:Beltrami}
%
The Beltrami pseudosphere is a bidimensional surface that we here choose to parameterize as
\begin{equation} \label{eq:Trumpetpar}
\begin{cases}
\;x\=L\;\exp\left(\frac{u}{R}\right)\;\cos\varphi\:;
\\[0.85ex]
\;y\=L\;\exp\left(\frac{u}{R}\right)\;\sin\varphi\:;
\\[-0.25ex]
\;z\=R\,\big(\arctanh f(u)-f(u)\big)\:,\qquad\qquad f(u)=\sqrt{1-\Big(\frac{L}{R}\,\exp\left(\frac{u}{R}\right)\Big)^2\:}\;.
\end{cases}
\end{equation}
%
%
As we can see by direct inspection, the parameterized Beltrami trumpet exists, in general, for $u\in\left[-\infty,\,R\:\log\left(\tfrac{R}{L}\right)\right]$.
The surface can be suitably embedded in $\mathbb{R}^3$ and it is well-defined over the whole non-singular part of the surface, the singular boundary being the maximal circle of radius $R$ corresponding to the limit value \,$\ustar=R\,\log\left(\tfrac{R}{L}\right)$.\par
\begin{figure}[H]
\captionsetup{skip=15pt,belowskip=-5pt,font=small,labelfont=small,format=hang}
\centering
\includegraphics[width=0.55\textwidth,keepaspectratio]{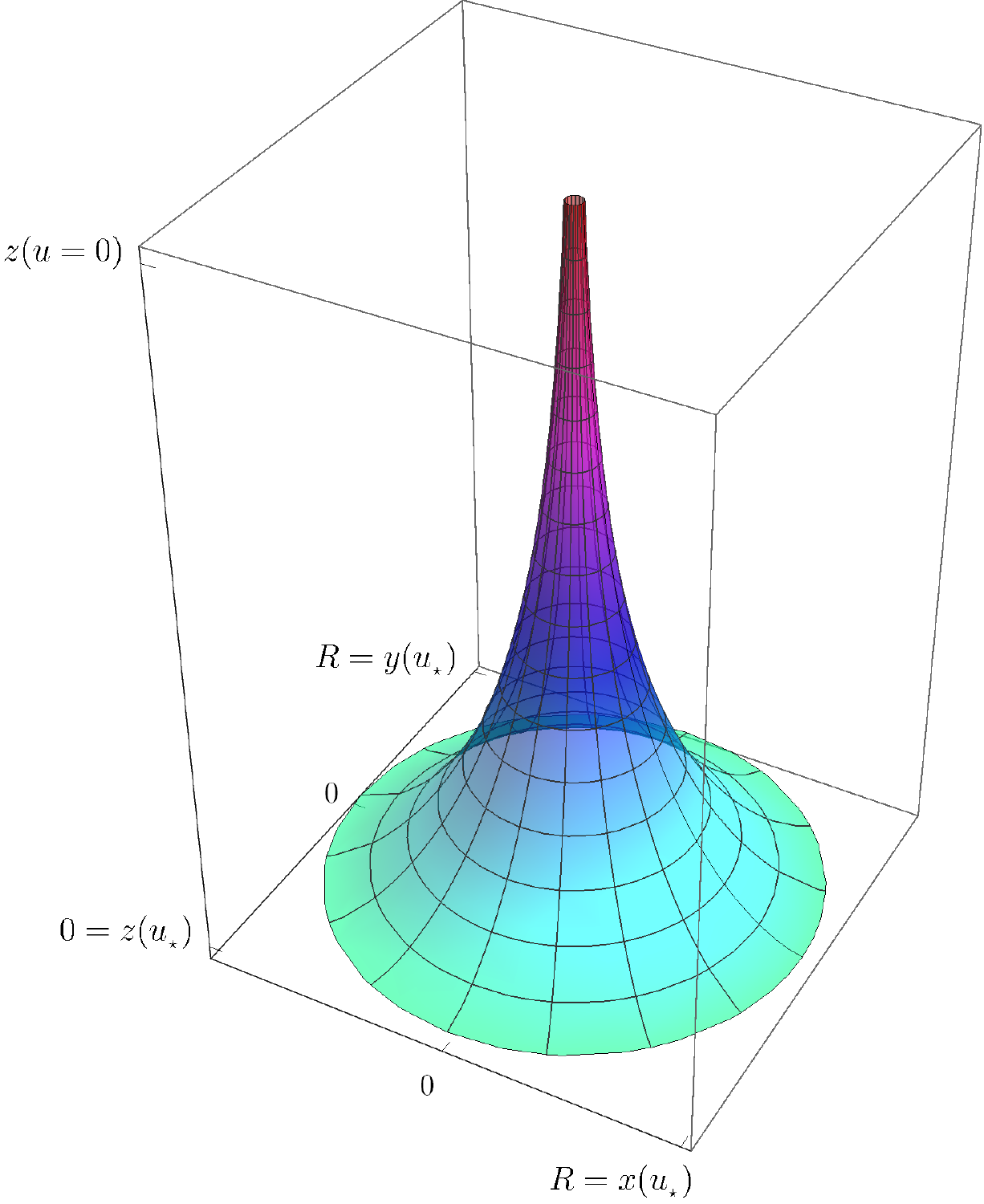}
\caption{Beltrami trumpet parametric plot, with $u\in\left[0,\,\ustar\right]$ and $v\in\left[0,\,2\pi\right]$.}
\label{fig:Trumpet}
\end{figure}
We can also notice that the above equations for the embedding are expressed in terms of the analogs of polar (or maybe better cylindrical) coordinates $u,\,\varphi$, so that we can easily move along the ``meridian'' and the ``parallel'' of the trumpet. Every coordinate is in fact expressed as a smooth, well-behaving, single-valued function (taking a full turn on a parallel has no effects on the values of $x,\,y$).\par
%

\paragraph{Graphene-like substrate.}
\sloppy
Even if the $u$--coordinate is unbounded from below (${u\to-\infty}$\, corresponding to \,${z\to\infty}$), when we consider a graphene-like membrane we have to take into account the considerations we made in Section \ref{sec:Intro} about the range of validity of the proposed charge carriers dynamics (not holding for too small radii), the continuum limit approximation and the difficulties in bending the surface beyond certain limits. All these aspects lead to the physical assumption of posing a limiting value for the surface (parallel) radius ${r=L\,e^{\tfrac{u}{R}}}$. Then, we should decide to make the natural assumption of minimum radius $r\geq L$, being $L$ much larger than the lattice length $\ell$ \:$\left(r\geq L\gg\ell\right)$. This, in turn, means that the $u$--coordinate is restricted to the interval ${u\in\left[0,\,\ustar\right]}$, with ${\ustar=R\:\log\left(\tfrac{R}{L}\right)}$.\par\smallskip
The metric on the pseudosphere \eqref{eq:Trumpetpar} has the form
\begin{equation}
\gmat=\left(
  \begin{array}{ccc}
    -1  &   0  &   0                           \\
    0   &   1  &   0                           \\
    0   &   0  &   L^2\:e^{\tfrac{2u}{R}}   \\
  \end{array}
\right)  \quad,
\end{equation}
so that the 1+2 spacetime line element reads
\begin{equation}
ds^2\=-\,dt^2\,+\,du^2\,+\,L^2\:e^{\tfrac{2u}{R}}\:d\varphi^2\:,
\end{equation}
the resulting Ricci scalar \,$\mathcal{R}=-\tfrac{2}{R^2}$\, being twice the Gaussian curvature $\mathcal{K}$\,.\par\smallskip
The vielbein can be easily found to be
\begin{equation}
\viel\=\left(
  \begin{array}{ccc}
    1  &  0  &  0                     \\
    0  &  1  &  0                     \\
    0  &  0  &  L\:e^{\tfrac{u}{R}} \\
  \end{array}
\right)  \quad,
\end{equation}
and it correctly satisfies the relation $\gmat=\viel\,\viel[\nu][b]\,\eta_{ab}$\,.
The curved gamma matrices explicitly read:
\begingroup%
\setlength{\belowdisplayshortskip}{8pt plus 1pt}%
\setlength{\belowdisplayskip}{10pt plus 1pt minus 1pt}%
\begin{equation}
\gamma_\mu\=\viel\,\gammafl_a
    \=\big(\,i\,\sigma_3\,,\;\sigma_1\,,\;L\,e^{\tfrac{u}{R}}\:\sigma_2\,\big)\;,
\label{eq:gammacurv}
\end{equation}
\endgroup
and the reader can verify that upper-indexed $\gamma^\mu$ satisfy the Clifford algebra $\left\{\gamma^\mu,\gamma^\nu\right\}=2\,\gmatinv\,\Id$\,.\par
Finally, the spin connection $\spc$ has non vanishing components
\begin{equation}
\spc[3][3][2]\=-\,\spc[3][2][3]\=\frac{L}{R}\,e^{\tfrac{u}{R}}\;,
\end{equation}
so that now we are able to explicitly express the Dirac equation \eqref{eq:Direqcurvmassless} in terms of the covariant derivative \eqref{eq:covder}.


\section{Analytic solutions in Beltrami geometry}\label{sec:Analyticsol}
Our challenge now is to obtain explicit $\psi$--solutions for the curved space Dirac fields satisfying \eqref{eq:Direqcurvmassless}, the corresponding relativistic pseudoparticles moving on a Beltrami surface.\par
The analytic solution of Dirac equation \eqref{eq:Direqcurvmassless} for charge carriers living on the Beltrami spacetime described in previous Subsect.\ \ref{subsec:Beltrami} has the explicit form
\begingroup%
\setlength{\belowdisplayshortskip}{3pt plus 1pt}%
\setlength{\belowdisplayskip}{5pt plus 1pt minus 1pt}%
\begin{equation}
\psi\=e^{-i\,\lambda\,E\,t}\;
        \left(\!\!
               \begin{array}{c}
               \phi_\textsc{a} \\[0.25ex]
               \phi_\textsc{b} \\
               \end{array}
        \!\!\right) \quad,
\label{eq:psisol}
\end{equation}
\endgroup
with
\begingroup%
\setlength{\abovedisplayshortskip}{3pt plus 1pt}%
\setlength{\abovedisplayskip}{5pt plus 1pt minus 1pt}
\begin{equation}\label{eq:phiABsol}
\begin{split}
\phi_\textsc{a}&\=\C\;e^{\,i\ml{\int}{du\,\xi(u)}}\;\,e^{i\,k\,\varphi}\;\frac{u}{R}\;e^{-\tfrac{u}{R}}\:,
\\[1ex]
\phi_\textsc{b}&\=\C\;e^{\,i\ml{\int}{du\,\xi(u)}}\;\,e^{i\,k\,\varphi}\;i\,\frac{u}{R}\;e^{-\tfrac{u}{R}}\;\,\B(u)\:,
\end{split}
\end{equation}
\endgroup
where $\lambda=\pm1$ labels states having positive/negative energy $E$, and $\C$ is a normalization constant that can be determined from the condition
\begin{equation}
\int d\Sigma\:\sqrt{g}\:\abs{\psi}^2\=1\;,
\label{eq:eq:normcond}
\end{equation}
being $\sqrt{g}=\sqrt{-\det(\gmat)}$\,.\; The functions $\xi(u)$ and $\B(u)$ are defined as
\begin{equation}
\begin{split}
\xi(u)&\=\left(\frac{1}{u}-\frac{k}{L}\,e^{-\tfrac{u}{R}}-\frac{1}{2R}\right)-\lambda\:E\:\B(u)\:,
\\[1ex]
\B(u)&\=\frac{\Iminus+\,\Iplus}{\Iminus-\,\Iplus}\:,
\end{split}
\end{equation}
expressed in terms of the modified Bessel function of the first kind%
\footnote{%
the modified Bessel function of first kind $I(n,Z)=Y$ is the function that satisfies the differential equation \;$Z^2\,Y^{\prime\prime}+Z\,Y^\prime-(Z^2+n^2)Y=0$\,; \,for certain arguments it has an explicit analytic expression, while it can always be evaluated to arbitrary numerical precision
}
\begin{equation}
I^{{}^\ms{(\pm)}}=I\left(\pm\frac12+i\,\lambda\,E\,R\;,\;\frac{kR}{L}\;e^{-\tfrac{u}{R}}\right)\:.
\end{equation}

\subsection{Experimental effects: local density of states}
We are now going to consider a simple experimental application for the above solution for pseudorelativistic charge carriers living in a 1+2 dimensional  curved background.\par
\begin{figure}[H]
\captionsetup{skip=10pt,belowskip=5pt,font=small,labelfont=small,format=hang}
\centering
\includegraphics[width=0.81\textwidth,keepaspectratio]{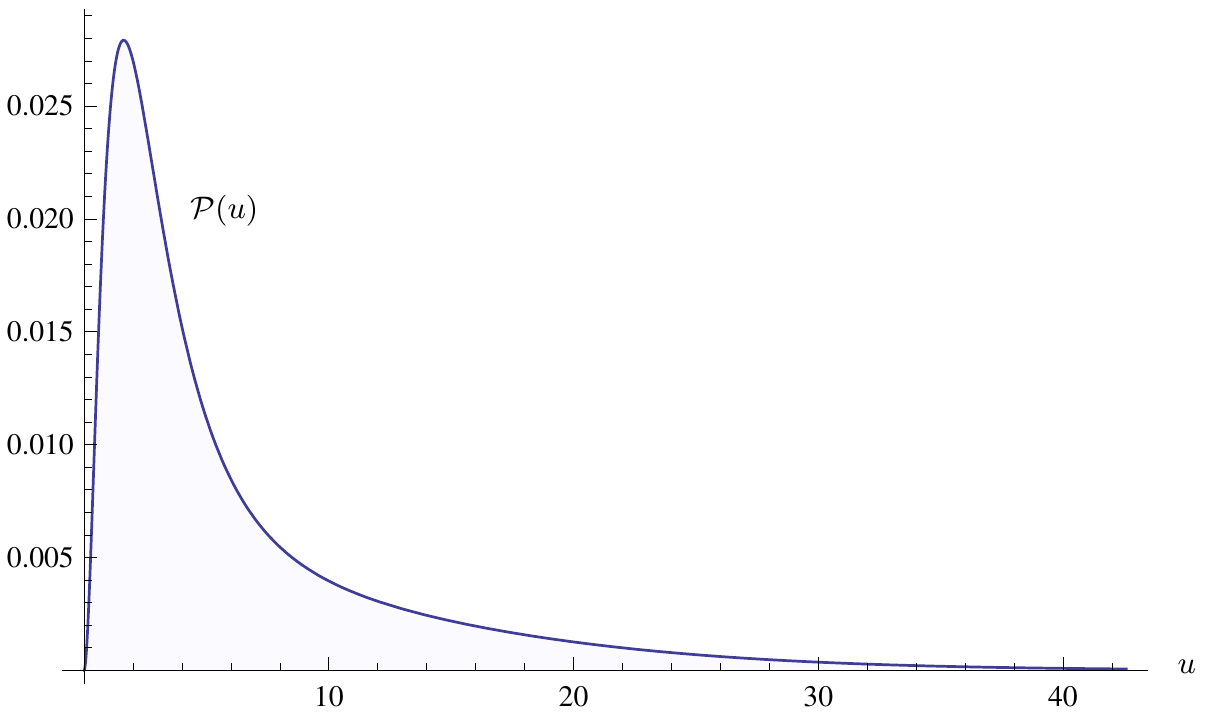}
\caption{Normalized probability density as a function of $u$ for a Beltrami-shaped sample surface with parameters \,${E=4\cdot10^{-4}}$, \,${\lambda=\pm1}$, \,${R=5}$, \,${L=0.001}$, \,${\mathcal{C}=1.415}$\,.}
\label{fig:PDu1}
\end{figure}

\begin{figure}[H]
\captionsetup{skip=10pt,belowskip=5pt,font=small,labelfont=small,format=hang}
\centering
\includegraphics[width=0.81\textwidth,keepaspectratio]{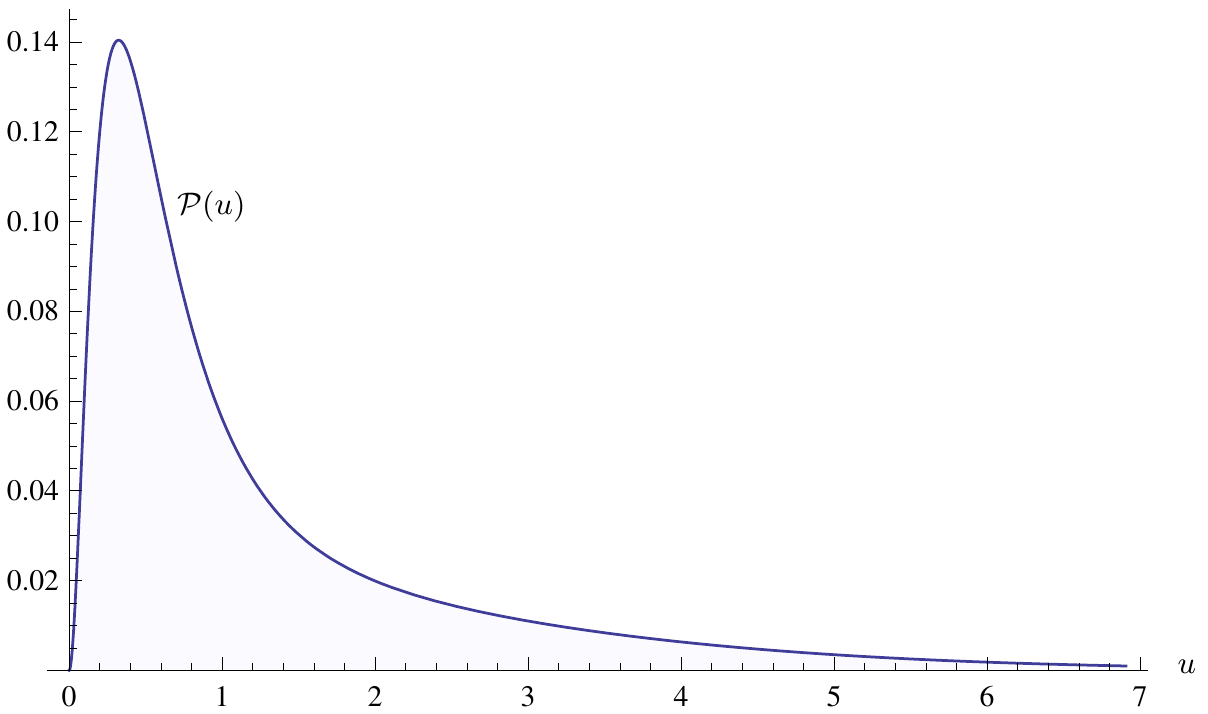}
\caption{Normalized probability density as a function of $u$ for a Beltrami-shaped sample surface with parameters \,${E=0.002}$, \,${\lambda=\pm1}$, \,${R=1}$, \,${L=0.001}$, \,${\mathcal{C}=3.174}$\,.}
\label{fig:PDu2}
\end{figure}

\begin{figure}[H]
\captionsetup{skip=10pt,belowskip=5pt,font=small,labelfont=small,format=hang}
\centering
\includegraphics[width=0.81\textwidth,keepaspectratio]{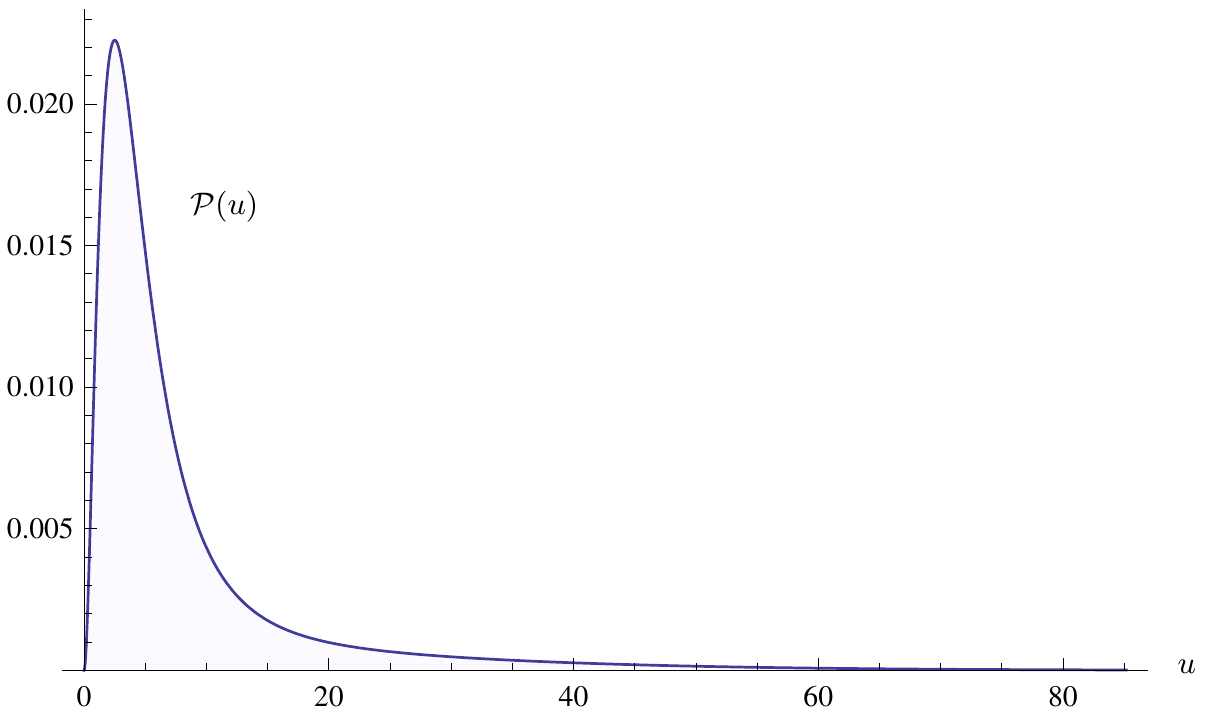}
\caption{Normalized probability density as a function of $u$ for a Beltrami-shaped sample surface with parameters \,${E=5\cdot10^{-4}}$, \,${\lambda=\pm1}$, \,${R=10}$, \,${L=0.002}$, \,${\mathcal{C}=0.455}$\,.}
\label{fig:PDu3}
\end{figure}

\smallskip
Since we have obtained the explicit Dirac solution $\psi$, we can consider the probability density $\PP=\sqrt{g}\:\abs{\psi}^2$, in terms of which the normalization condition \eqref{eq:eq:normcond} in Beltrami geometry can be explicitly written as
\begin{equation}
\int d\Sigma\;\PP(u,\varphi)
    \=2\pi\int{\!du\:L\:e^{\tfrac{u}{R}}\:\abs{\psi(u,\varphi)}^2}\=1\;.
\end{equation}
Using now our new solutions \eqref{eq:psisol}, \eqref{eq:phiABsol} for the Beltrami surface, it is possible to obtain the properly normalized probability density, as shown in \cref{fig:PDu1,fig:PDu2,fig:PDu3}.
We should use the appropriate precautions discussed in Sections \ref{sec:Intro} and \ref{subsec:Beltrami}, that is considering the proper interval of variation for the $u$--coordinate in order to satisfy the correct continuum limit for the membrane.
\medskip
As we can appreciate, for the chosen parameters, the peak is localized in correspondence of small values for $u$, the charge carriers being then mainly localized on the throat of the trumpet.\par\smallskip
For a given a location $\X$ on the layer surface and energy value $E$, the local density of states (LDOS) of the sample can be defined as \cite{Chen1993introduction}
\begin{equation}
\LDOS(E,\X,0)\=\frac{1}{\varepsilon}\:\sum_{E-\varepsilon}^{E}\PP(\X)\:,
\label{eq:LDOS}
\end{equation}
%
for sufficiently small values of $\varepsilon$, where the $0$--coordinate means that we are considering pseudoparticles on the sample surface (i.e.\ zero distance from the substrate surface). The physical meaning of the LDOS for our bidimensional sample is the number of charge carriers per unit surface and unit energy range of size $\varepsilon$, at a given surface location $\X$ and energy $E$. The sample LDOS is not only an interesting direct observable for the predicted quantum behaviour, but also a substrate feature of immense importance for electronic applications, being the availability of empty valence and conduction states (states below and above the Fermi level) crucial for the transition rates.

\paragraph{Measurements.}
The sample LDOS can be detected using a scanning tunneling microscope (STM). The latter is an experimental device based on quantum mechanical tunneling, in which the wave-like properties of charge carries allow them to penetrate through a potential barrier, into regions that are forbidden to them in the classical picture. STM spectroscopy provides insight into the surface electronic properties of the substrate, being the tunneling current strongly affected by the local density of states $\LDOS$. The latter is in turn related to the probability density $\PP$ through definition \eqref{eq:LDOS}.\par
%
A typical STM device consists of a very sharp conductive tip which is brought within tunneling distance ($<\mathrm{nm}$) from a sample surface, using a three-dimensional piezoelectric scanner.
Let us imagine that electrons fill energy levels up to the Fermi level -- which defines an upper boundary similar to the sea level -- above which we find activated charge carriers. The Fermi level of a material can be raised/lowered with respect to a second material by applying an appropriate voltage. Thus, to obtain a tunneling current through the gap between the sample and the tip, a suitable bias voltage $\V$ can be applied, causing charge carriers to tunnel across the gap%
\footnote{%
to simplify our discussion, we are assuming that both materials have the same Fermi level}.
In particular, when a negative bias is applied to the sample, its Fermi level is raised and the charge carriers of the filled occupied states can tunnel into the unoccupied state of the tip, while the opposite occurs for a positive bias (filling of the sample empty states).\par
%
We are interested in the STM--map setup, where the density of states at some fixed energy, is mapped as a function of the position $(u,\varphi)$ on the sample surface. Let us assume $\varepsilon=e\,\V$ to be very small with respect to the work function $\Phi_\textsc{w}$ (minimum energy required to extract an electron from the surface), so that the sample states with energy lying between $\EF-\varepsilon$ and $\EF$ are very close to the Fermi level and have non-zero probability of tunneling into the tip.
The resulting tunneling current $\I$ is directly proportional to the number of states on the substrate within our energy range of width $\varepsilon$, this number depending on the local properties of the surface. Including all the sample states in the chosen energy range, the measured tunneling current can be modelled, in first approximation, as \cite{Chen1993introduction}
\begin{equation}
\I~\propto\,\sum_{\EF-e\,\V}^{\EF}\PP\;e^{-2\,\kappa\,d}\;,
\end{equation}
where $\kappa$ is some decay constant in the barrier separation depending on $\Phi_\textsc{w}$\,. The exponential function gives the suppression for charge carriers tunneling in the classically forbidden region of width $d$ (sample-tip separation).
The tunneling current can be then measured, for constant separation $d$, at different $\X$ positions and, for sufficiently small $\V$, it can be conveniently expressed in terms of the LDOS of the sample as \cite{Chen1993introduction}:
\begin{equation}
\I(\X)~\propto~\,\LDOS(\EF,\X,0)\;e^{-2\,\kappa\,d}\;e\,\V\;.
\end{equation}
In summary, tunneling current measured by STM mapping, at small bias voltage $\V$ and fixed tip-sample separation $d$, is proportional to the local density of states of the sample. 
In particular, we could scan our Beltrami surface at different positions varying the $u$--coordinate, mapping inhomogeneities in the local density of states \,$\LDOS(\EF,u,0)$, that in turn depend on the predicted surface probability density $\PP(u)$.\par
%
STM can operate in ambient atmosphere as well as in high vacuum; when a high-vacuum configuration is employed, its purpose is not to improve the performance of the STM but rather to ensure the cleanliness of the sample surface. We also remark that the LDOS obtained with a STM is not limited by the position of the Fermi energy, since both occupied and empty states are accessible \cite{Andrei:2012my}.
For finite bias voltage and different Fermi levels for the sample and tip material, the functional form of the tunneling current and its relation with the sample local density of states can be easily obtained from Bardeen time-dependent perturbation approach \cite{Chen1993introduction}.

\section{Conclusions}
A deeper intertwining of different scientific areas can really provide an important step forward in our understanding of various, fundamental physical aspects of our world. In particular, it has been shown that the considerable gap between high energy physics and condensed matter -- due to mutually independent mathematical formulation and developments -- can be reduced using a multidisciplinary approach (see e.g.\ \cite{Maldacena:1997re,Witten:1998qj,Zurek:1996sj,Volovik:2000ua,Ummarino:2017bvz,Ummarino:2019cvw,Ummarino2020josephson,Gallerati2020interaction,Sepehri:2016nuv,Capozziello:2018mqy,Capozziello:2020ncr,Baeuerle:1996zz,Ruutu:1995qz,novello2002artificial,Barcelo:2005fc,Andrianopoli:2019sip,Gallerati2020supersymmetric}).
Following this spirit, we have used along the paper different techniques from high energy physics, differential geometry and general relativity, applying them to the study of pseudoparticles living in a curved, real bidimensional surface.
\par
When dealing with graphene-like materials, one immediately realizes the need for a correct quantum relativistic field description, these special materials implementing the physics of Dirac relativistic fermions in a real condensate system. As we have discussed, we also gain the concrete possibility to observe the analogues of gravity effects, driven by the specific curved spacetime, dictated in turn by the chosen membrane structure. In this work, in particular, we have found an analytic, explicit expression for the pseudorelativistic charge carriers modes in the notable framework of the Beltrami geometry, being subsequently able to characterize a simple observable like the pseudoparticles local density of states.
Clearly, the same approach finds application also in different spacetime geometries; one example is given by  the BTZ geometry \cite{Banados:1992wn}, that under certain conditions can mimic the background for low-energy electron excitations of a curved graphene sheet \cite{Cvetic:2012vg,Kandemir:2019zyt}.

\bigskip

\section*{\normalsize Acknowledgments}
\vspace{-2.5pt}
I would like to thank professor G.\ A.\ Ummarino for extremely helpful discussions during the preparation of this report.
I would also like to thank prof.\ M.\ Trigiante, prof.\ F.\ Laviano and prof.\ A.\ Carbone
that supported these studies with their funds.\par\bigskip
%


\newpage
\appendix
\addtocontents{toc}{\protect\setcounter{tocdepth}{1}}
\addtocontents{toc}{\protect\addvspace{3.5pt}}%
\numberwithin{equation}{section}%
\numberwithin{figure}{section}%

\section{Conventions}\label{app:Conventions}
\paragraph{Dirac equation.}
The Dirac equation in Minkowski spacetime is the result of the construction of a relativistic field equation, whose squared wave function modulus could be consistently interpreted as a probability density. To satisfy these conditions, the equation is of first order in time-derivative, while relativistic invariance requires the equation to be first order in space-derivatives too. The final explicit form must fulfill the requests of Lorentz covariance and satisfy the Klein-Gordon equation, and reads
\begin{equation}
(i\,\gammafl^a \dd[a] - m\,\Id)\;\psi(x)\=0\;,
\label{eq:Diraceqflat}
\end{equation}
together with the condition
\begin{equation}
\left\{\gammafl^a,\,\gammafl^b\right\}\=2\,\eta^{ab}\,\Id\:,
\label{eq:Cliffalgflat}
\end{equation}
that is usually referred to as Clifford algebra, the matrix $\eta^{ab}$ being the inverse of the Minkowski flat metric $\eta_{ab}\,$.\par
For the sake of notational simplicity, in eq.\ \eqref{eq:Diraceqflat} we have omitted the spinorial indices of $\psi\equiv\psi^\beta$\, and \,$\gammafl^a={(\gammafl^a)^\alpha}_\beta$\,. Spinors are objects that transform as scalars under general space-time coordinate transformations, while they trasform in a spinor representation $\mathscr{R}$ under the local Lorentz group:
\begingroup
\setlength{\abovedisplayshortskip}{2pt plus 1pt}%
\setlength{\abovedisplayskip}{3pt plus 1pt minus 2pt}
\setlength{\belowdisplayshortskip}{5pt plus 1pt}%
\setlength{\belowdisplayskip}{6pt plus 2pt minus 2pt}%
\begin{equation}
\psi^{\prime\,\alpha}(x) \=
{\mathscr{R}\big[\Lambda(x)\big]}^{{\!}^\ml{\alpha}}{}_{{\!}_\ml{\beta}}\;\psi^\beta(x)\:.
\end{equation}
\endgroup
Using the explicit form of the Lorentz generators to construct the Pauli-Lubanski operator, it can be easily shown that the particle has spin $s=\tfrac12$\,.

\subsection{Curved spaces} \label{subapp:curvspace}
Einstein's theory of gravitation relies on the symmetry principle of invariance under general coordinate transformations, that, in turn, can be viewed as local spacetime transformations generated by the local translation generators. The gravitational force can be then geometrically modelled in terms of the spacetime curvature.\par
In order to conveniently describe general relativity scenarios together with spinorial fields, one should introduce some tools to describe transformation rules generalized to curved backgrounds, leading to the so-called vielbein formalism.
\paragraph{Vielbein formalism.}
Let us consider a set of coordinates that is locally inertial, so that one can apply the usual Lorentz spinor behaviour, and imagine to find a way to translate back to the original coordinate frame. More precisely, let $y^a(x_0)$ denote a coordinate frame that is inertial at the space-time point $x_0$: we shall call these the ``Lorentz'' coordinates. Then,
\begin{equation}
\viel(x) \= \frac{\partial y^a (x_0)}{\partial x^\mu}
\end{equation}
gives the so-called \emph{vielbein}: it defines a local set of tangent frames of the spacetime manifold and, under general coordinate transformations, it transforms covariantly as
\begin{equation}
{e'_\mu}^{\!a}(x') \=
\frac{\partial x^\nu}{\partial x'^\mu}\:\,\viel[\nu][a](x)\;,
\end{equation}
while a Lorentz transformation leads to
\begin{equation}
{e'_\mu}^a (x) \= {\Lambda^{{\!}^\ml{a}}}_b\:\,\viel[\mu][b](x)\;.
\end{equation}
The space-time metric, in particular, can be expressed as
\begin{equation}
g_{\mu\nu}(x) \= \viel[\mu][a](x)\;\viel[\nu][b](x)\;\eta_{ab}\;,
\end{equation}
in terms of the Minkowski flat metric $\eta_{ab}$\,.
The original constant $\gammafl_a$ matrices of the inertial frame can be converted into the new $\gamma_\mu$ matrices of the curved background by the action of the vielbein:
\begin{equation}
\gamma_\mu(x)\=\viel(x)\:\gammafl_a\;,
\end{equation}
while the inverse vielbein $\vielinv$ performs the transformation in the other direction. The vielbein thus takes Lorentz (flat) latin indices to coordinate basis (curved) greek indices. The gamma matrices with upper indices
\begin{equation}
\gamma^\mu\=\gmatinv\,\gamma_\nu\;,
\end{equation}
satisfy the relation:
\begin{equation}
\{\gamma^\mu,\,\gamma^\nu\}\=2\,g^{\mu\nu}\,\Id\;,
\end{equation}
that holds in curved backgrounds and is the equivalent form of the previous, flat Clifford algebra \eqref{eq:Cliffalgflat}.

\paragraph{Covariant derivative, spin connection.}
The choice of the locally inertial frame $y^a$ is defined up to Lorentz transformations given by the Lorentz generators $M_{ab}$\,. In order to couple fields, we define the covariant derivatives:
\begin{equation}
\mathcal{D}_\mu \= \dd + \frac14\,\spc\,M_{ab}\;,
\end{equation}
where
\begin{equation}
M_{ab} \= \frac12\,\left[\gammafl_a,\gammafl_b\right]\;.
\end{equation}
The $\spc$ object defines the \emph{spin connection}, that can be seen as the gauge field of the local Lorentz group, the corresponding field strength given by the Riemann curvature tensor, and is determined through the vielbein postulate (tetrad covariantly constant) \cite{Kleinert1989gauge,Katanaev:1992kh}:
\begin{equation}
\mathcal{D}_\mu \viel[\nu] - \Conn\,\viel[\lambda] \= 0 \;.
\end{equation}
The latter is written in terms of the affine connection $\Conn$
\begin{equation}
\Conn \= \frac12\,\gmatinv[\sigma][\lambda]
       \left(\dd\gmat[\nu][\sigma]+\dd[\nu]\gmat[\mu][\sigma]-\dd[\sigma]\gmat\right)\;.
\end{equation}
while the explicitly expression for the spin connection is found to be
\begin{equation}
\spc \= \viel[\nu][a]\,\dd\vieluu[\nu][b]+\viel[\nu][a]\,\Conn[\mu][\lambda][\nu]\,\vieluu[\lambda][b]\;.
\end{equation}
Finally, the Dirac equation in curved spacetime can be written as:
\begin{equation}
\left(i\,\gamma^\mu\,\DD - m\,\Id\right)\;\psi \= 0 \;.
\label{miao}
\end{equation}


\newpage

\hypersetup{linkcolor=blue}
\phantomsection 
\addtocontents{toc}{\protect\addvspace{4.5pt}}
\addcontentsline{toc}{section}{References} 
\bibliographystyle{mybibstyle}
\bibliography{bibliografia} 

\providecommand{\href}[2]{#2}\begingroup\begin{thebibliography}{10}

\bibitem{katsnelson2007graphene}
M.I. Katsnelson and K.S. Novoselov, \textit{``Graphene: New bridge between
  condensed matter physics and quantum electrodynamics''}, Solid State
  Communications \textbf{143} (2007), n.~1, 3--13.

\bibitem{geim2007rise}
Andre~K. Geim and Konstantin~S. Novoselov, \textit{``The rise of graphene''},
  Nature materials \textbf{6} (2007), n.~3, 183.

\bibitem{CastroNeto:2009zz}
A.H. Castro~Neto, F.~Guinea, N.M.R. Peres, K.S. Novoselov and A.K. Geim,
  \textit{``{The electronic properties of graphene}''}, Rev. Mod. Phys.
  \textbf{81} (2009) 109--162,
  [\href{http://arxiv.org/abs/0709.1163}{\texttt{arXiv:0709.1163}}].

\bibitem{Semenoff:1984dq}
Gordon~W. Semenoff, \textit{``{Condensed Matter Simulation of a
  Three-dimensional Anomaly}''}, Phys.\ Rev.\ Lett. \textbf{53} (1984) 2449.

\bibitem{novoselov2004electric}
K.S. Novoselov, A.K. Geim, S.V. Morozov, D.~Jiang, Y.~Zhang, S.V. Dubonos, I.V.
  Grigorieva and A.A. Firsov, \textit{``Electric field effect in atomically
  thin carbon films''}, Science \textbf{306} (2004), n.~5696, 666--669.

\bibitem{novoselov2005twodimato}
K.S. Novoselov, D.~Jiang, F.~Schedin, T.J. Booth, V.V. Khotkevich, S.V. Morozov
  and A.K. Geim, \textit{``Two-dimensional atomic crystals''}, Proc.\ of the
  National Academy of Sciences of the USA \textbf{102} (2005), n.~30,
  10451--10453.

\bibitem{Novoselov:2005kj}
K.~S. Novoselov, A.~K. Geim, S.~V. Morozov, D.~Jiang, M.~I. Katsnelson, I.~V.
  Grigorieva, S.~V. Dubonos and A.~A. Firsov, \textit{``{Two-dimensional gas of
  massless Dirac fermions in graphene}''}, Nature \textbf{438} (2005) 197,
  [\href{http://arxiv.org/abs/cond-mat/0509330}{\texttt{cond-mat/0509330}}].

\bibitem{Gusynin:2006ym}
V.~P. Gusynin, S.~G. Sharapov and J.~P. Carbotte, \textit{``{Unusual microwave
  response of Dirac quasiparticles in graphene}''}, Phys. Rev. Lett.
  \textbf{96} (2006) 256802,
  [\href{http://arxiv.org/abs/cond-mat/0603267}{\texttt{cond-mat/0603267}}].

\bibitem{Birrell:1982ix}
N.D. Birrell and P.C.W. Davies, \textit{``{Quantum Fields in Curved Space}''};
  Cambridge Univ. Press, Cambridge, UK (1984).

\bibitem{Brill:1966tia}
Dieter~R. Brill and J.M. Cohen, \textit{``{Cartan Frames and the General
  Relativistic Dirac Equation}''}, J.\ Math.\ Phys. \textbf{7} (1966), n.~2,
  238.

\bibitem{Wald:1984rg}
Robert~M. Wald, \textit{``{General Relativity}''}; Chicago Univ. Pr., Chicago,
  USA (1984).

\bibitem{Cortijo:2006mh}
Alberto Cortijo and Maria A.~H. Vozmediano, \textit{``{Electronic properties of
  curved graphene sheets}''}, EPL \textbf{77} (2007), n.~4, 47002,
  [\href{http://arxiv.org/abs/cond-mat/0603717}{\texttt{cond-mat/0603717}}].

\bibitem{Cortijo:2006xs}
Alberto Cortijo and Maria A.~H. Vozmediano, \textit{``{Effects of topological
  defects and local curvature on the electronic properties of planar
  graphene}''}, Nucl. Phys. \textbf{B763} (2007) 293--308,
  [\href{http://arxiv.org/abs/cond-mat/0612374}{\texttt{cond-mat/0612374}}].

\bibitem{Vozmediano:2010zz}
M.A.H. Vozmediano, M.I. Katsnelson and F.~Guinea, \textit{``{Gauge fields in
  graphene}''}, Phys. Rept. \textbf{496} (2010) 109--148,
  [\href{http://arxiv.org/abs/1003.5179}{\texttt{arXiv:1003.5179}}].

\bibitem{Gorbar:2007kd}
E.V. Gorbar and V.P. Gusynin, \textit{``{Gap generation for Dirac fermions on
  Lobachevsky plane in a magnetic field}''}, Annals Phys. \textbf{323} (2008)
  2132--2146,
  [\href{http://arxiv.org/abs/0710.2292}{\texttt{arXiv:0710.2292}}].

\bibitem{Boada:2010sh}
O.~Boada, A.~Celi, J.~I. Latorre and M.~Lewenstein, \textit{``{Dirac Equation
  For Cold Atoms In Artificial Curved Spacetimes}''}, New J. Phys. \textbf{13}
  (2011) 035002,
  [\href{http://arxiv.org/abs/1010.1716}{\texttt{arXiv:1010.1716}}].

\bibitem{Gallerati:2018dgm}
Antonio Gallerati, \textit{``{Graphene properties from curved space Dirac
  equation}''}, Eur.\ Phys.\ J.\ Plus \textbf{134} (2019), n.~5, 202,
  [\href{http://arxiv.org/abs/1808.01187}{\texttt{arXiv:1808.01187}}].

\bibitem{Kotov:2010yh}
Valeri~N. Kotov, Bruno Uchoa, Vitor~M. Pereira, A.H.~Castro Neto and F.~Guinea,
  \textit{``{Electron-Electron Interactions in Graphene: Current Status and
  Perspectives}''}, Rev. Mod. Phys. \textbf{84} (2012) 1067,
  [\href{http://arxiv.org/abs/1012.3484}{\texttt{arXiv:1012.3484}}].

\bibitem{novello2002artificial}
Mario Novello, Matt Visser and Grigory~E. Volovik, \textit{``Artificial black
  holes''}; World Scientific (2002).

\bibitem{Barcelo:2005fc}
Carlos Barcelo, Stefano Liberati and Matt Visser, \textit{``{Analogue
  gravity}''}, Living Rev. Rel. \textbf{8} (2005) 12.

\bibitem{Alvarez:2011gd}
Pedro~D. Alvarez, Mauricio Valenzuela and Jorge Zanelli,
  \textit{``{Supersymmetry of a different kind}''}, JHEP \textbf{04} (2012)
  058, [\href{http://arxiv.org/abs/1109.3944}{\texttt{arXiv:1109.3944}}].

\bibitem{Alvarez:2013tga}
Pedro~D. Alvarez, Pablo Pais and Jorge Zanelli, \textit{``{Unconventional
  supersymmetry and its breaking}''}, Phys. Lett. \textbf{B735} (2014)
  314--321, [\href{http://arxiv.org/abs/1306.1247}{\texttt{arXiv:1306.1247}}].

\bibitem{Andrianopoli:2019sip}
L.~Andrianopoli, B.~L. Cerchiai, R.~D'Auria, A.~Gallerati, R.~Noris,
  M.~Trigiante and J.~Zanelli, \textit{``{$\mathcal{N}$-extended $D = 4$
  supergravity, unconventional SUSY and graphene}''}, JHEP \textbf{01} (2020)
  084, [\href{http://arxiv.org/abs/1910.03508}{\texttt{arXiv:1910.03508}}].

\bibitem{Gallerati2020supersymmetric}
Antonio Gallerati, \textit{``{Supersymmetric theories and graphene}''}, PoS
  ICHEP 2020 (2021).

\bibitem{osipov2005electronic}
V.A. Osipov and D.V. Kolesnikov, \textit{``Electronic properties of curved
  carbon nanostructures''}, Romanian Journal of Physics \textbf{50} (2005),
  n.~3/4, 457.

\bibitem{kolesnikov2006continuum}
D.V.\ Kolesnikov and V.A.\ Osipov, \textit{``The continuum gauge field-theory
  model for low-energy electronic states of icosahedral fullerenes''}, Eur.
  Phys. J. \textbf{B49} (2006), n.~4, 465--470.

\bibitem{morpurgo2006intervalley}
A.F. Morpurgo and F.~Guinea, \textit{``Intervalley scattering, long-range
  disorder, and effective time-reversal symmetry breaking in graphene''},
  Physical Review Letters \textbf{97} (2006), n.~19, 196804.

\bibitem{lee2009surface}
Dung-Hai Lee, \textit{``{Surface states of topological insulators: The Dirac
  fermion in curved two-dimensional spaces}''}, Physical Review Letters
  \textbf{103} (2009), n.~19, 196804.

\bibitem{Beltrami2017saggio}
E.~Beltrami, \textit{``{Saggio di interpretazione della Geometria
  non-euclidea}''}, Giornale di matematiche ad uso degli studenti delle
  università italiane \textbf{6} (1868) 284--312.

\bibitem{Iorio:2011yz}
Alfredo Iorio and Gaetano Lambiase, \textit{``{The Hawking-Unruh phenomenon on
  graphene}''}, Phys. Lett. \textbf{B716} (2012) 334--337,
  [\href{http://arxiv.org/abs/1108.2340}{\texttt{arXiv:1108.2340}}].

\bibitem{Chen:2012uc}
Pisin Chen and Haret Rosu, \textit{``{Note on Hawking-Unruh effects in
  graphene}''}, Mod. Phys. Lett. \textbf{A27} (2012) 1250218,
  [\href{http://arxiv.org/abs/1205.4039}{\texttt{arXiv:1205.4039}}].

\bibitem{Hawking:1974rv}
S.~W. Hawking, \textit{``{Black hole explosions}''}, Nature \textbf{248} (1974)
  30--31.

\bibitem{Unruh:1976db}
W.~G. Unruh, \textit{``{Notes on black hole evaporation}''}, Phys. Rev.
  \textbf{D14} (1976) 870.

\bibitem{Iorio:2012xg}
Alfredo Iorio, \textit{``{Using Weyl symmetry to make Graphene a real lab for
  fundamental physics}''}, Eur.\ Phys.\ J.\ Plus \textbf{127} (2012) 156,
  [\href{http://arxiv.org/abs/1207.6929}{\texttt{arXiv:1207.6929}}].

\bibitem{Iorio:2013ifa}
Alfredo Iorio and Gaetano Lambiase, \textit{``{Quantum field theory in curved
  graphene spacetimes, Lobachevsky geometry, Weyl symmetry, Hawking effect, and
  all that}''}, Phys. Rev. \textbf{D90} (2014), n.~2, 025006,
  [\href{http://arxiv.org/abs/1308.0265}{\texttt{arXiv:1308.0265}}].

\bibitem{morresi2020exploring}
Tommaso Morresi, Daniele Binosi, Stefano Simonucci, Riccardo Piergallini,
  Stephan Roche, Nicola~M Pugno and Taioli Simone, \textit{``Exploring event
  horizons and hawking radiation through deformed graphene membranes''}, 2D
  Materials \textbf{7} (2020), n.~4, 041006.

\bibitem{Wang2015rare}
Jinying Wang, Shibin Deng, Zhongfan Liu and Zhirong Liu, \textit{``{The rare
  two-dimensional materials with Dirac cones}''}, National Science Review
  \textbf{2} (Jan., 2015) 22--39.

\bibitem{Cahangirov2009twoandone}
S.~Cahangirov, M.~Topsakal, E.~Akt\"{u}rk, H.~{\c{S}}ahin and S.~Ciraci,
  \textit{``{Two- and One-Dimensional Honeycomb Structures of Silicon and
  Germanium}''}, Phys. Rev. Lett. \textbf{102} (June, 2009).

\bibitem{Malko2012competition}
Daniel Malko, Christian Neiss, Francesc Vi{\~{n}}es and Andreas G\"{o}rling,
  \textit{``{Competition for Graphene: Graphynes with Direction-Dependent Dirac
  Cones}''}, Phys. Rev. Lett. \textbf{108} (Feb., 2012).

\bibitem{Xu2014twodim}
Li-Chun Xu, Ru-Zhi Wang, Mao-Sheng Miao, Xiao-Lin Wei, Yuan-Ping Chen, Hui Yan,
  Woon-Ming Lau, Li-Min Liu and Yan-Ming Ma, \textit{``{Two dimensional Dirac
  carbon allotropes from graphene}''}, Nanoscale \textbf{6} (2014), n.~2,
  1113--1118.

\bibitem{Zhou2014semimetallic}
Xiang-Feng Zhou, Xiao Dong, Artem~R. Oganov, Qiang Zhu, Yongjun Tian and
  Hui-Tian Wang, \textit{``{Semimetallic Two-Dimensional Boron Allotrope with
  Massless Dirac Fermions}''}, Phys. Rev. Lett. \textbf{112} (Feb., 2014).

\bibitem{Pardo2009half}
Victor Pardo and Warren~E. Pickett, \textit{``{Half-Metallic Semi-Dirac-Point
  Generated by Quantum Confinement in TiO$_2$/VO$_2$ Nanostructures}''}, Phys.
  Rev. Lett. \textbf{102} (Apr., 2009).

\bibitem{Katayama2006pressure}
Shinya Katayama, Akito Kobayashi and Yoshikazu Suzumura,
  \textit{``{Pressure-Induced Zero-Gap Semiconducting State in Organic
  Conductor $\alpha$-({BEDT}-{TTF})2I3Salt}''}, Journal of the Physical Society
  of Japan \textbf{75} (May, 2006) 054705.

\bibitem{Li2014gapless}
Weifeng Li, Meng Guo, Gang Zhang and Yong-Wei Zhang, \textit{``{Gapless MoS$_2$
  allotrope possessing both massless Dirac and heavy fermions}''}, Phys. Rev. B
  \textbf{89} (May, 2014).

\bibitem{Zhu2007simulation}
Shi-Liang Zhu, Baigeng Wang and L.-M. Duan, \textit{``{Simulation and Detection
  of Dirac Fermions with Cold Atoms in an Optical Lattice}''}, Phys. Rev. Lett.
  \textbf{98} (June, 2007).

\bibitem{Peres:2010mx}
N.~M.~R. Peres, \textit{``{Colloquium: The Transport properties of graphene: An
  Introduction}''}, Rev. Mod. Phys. \textbf{82} (2010) 2673--2700,
  [\href{http://arxiv.org/abs/1007.2849}{\texttt{arXiv:1007.2849}}].

\bibitem{deJuan:2012hxm}
Fernando de~Juan, Mauricio Sturla and Maria A.~H. Vozmediano, \textit{``{Space
  dependent Fermi velocity in strained graphene}''}, Phys. Rev. Lett.
  \textbf{108} (2012), n.~22, 227205,
  [\href{http://arxiv.org/abs/1201.2656}{\texttt{arXiv:1201.2656}}].

\bibitem{Kleinert1989gauge}
H.~Kleinert, \textit{``{Gauge fields in condensed matter}''}; Singapore, World
  Scientific (1989).

\bibitem{Katanaev:1992kh}
M.O. Katanaev and I.V. Volovich, \textit{``{Theory of defects in solids and
  three-dimensional gravity}''}, Annals Phys. \textbf{216} (1992) 1--28.

\bibitem{Guinea2008midgap}
F.~Guinea, M.I. Katsnelson and M.A.H. Vozmediano, \textit{``Midgap states and
  charge inhomogeneities in corrugated graphene''}, Physical Review B
  \textbf{77} (Feb., 2008).

\bibitem{Levy2010strain}
N.~Levy, S.~A. Burke, K.~L. Meaker, M.~Panlasigui, A.~Zettl, F.~Guinea,
  A.~H.~C. Neto and M.~F. Crommie, \textit{``{Strain-Induced Pseudo-Magnetic
  Fields Greater Than 300 Tesla in Graphene Nanobubbles}''}, Science
  \textbf{329} (July, 2010) 544--547.

\bibitem{Morozov2006strong}
Sergei~V. Morozov, Kostya~S. Novoselov, M.I. Katsnelson, F.~Schedin, L.A.
  Ponomarenko, D.~Jiang and Andre~K. Geim, \textit{``Strong suppression of weak
  localization in graphene''}, Physical review letters \textbf{97} (2006),
  n.~1, 016801.

\bibitem{Stegmann:2015mjp}
Thomas Stegmann and Nikodem Szpak, \textit{``{Current flow paths in deformed
  graphene: from quantum transport to classical trajectories in curved
  space}''}, New J. Phys. \textbf{18} (2016), n.~5, 053016,
  [\href{http://arxiv.org/abs/1512.06750}{\texttt{arXiv:1512.06750}}].

\bibitem{Vozmediano:2008zz}
Maria~A.H. Vozmediano, Fernando de~Juan and Alberto Cortijo, \textit{``{Gauge
  fields and curvature in graphene}''}, J. Phys. Conf. Ser. \textbf{129} (2008)
  012001, [\href{http://arxiv.org/abs/0807.3909}{\texttt{arXiv:0807.3909}}].

\bibitem{deJuan2007charge}
Fernando de~Juan, Alberto Cortijo and Maria A.~H. Vozmediano, \textit{``Charge
  inhomogeneities due to smooth ripples in graphene sheets''}, Physical Review
  B \textbf{76} (Oct., 2007).

\bibitem{Carpio2008dislocations}
Ana Carpio, Luis~L.\ Bonilla, Fernando de~Juan and Maria~A.H.\ Vozmediano,
  \textit{``Dislocations in graphene''}, New J. Phys. \textbf{10} (2008), n.~5,
  053021.

\bibitem{deJuan2010dislocations}
Fernando de~Juan, Alberto Cortijo and Maria~A.H. Vozmediano,
  \textit{``Dislocations and torsion in graphene and related systems''},
  Nuclear physics B \textbf{828} (2010), n.~3, 625--637.

\bibitem{seung1988defects}
Hyunjune~Sebastian Seung and David~R. Nelson, \textit{``Defects in flexible
  membranes with crystalline order''}, Physical Review A \textbf{38} (1988),
  n.~2, 1005.

\bibitem{Gupta:2008mn}
Kumar~S. Gupta and Siddhartha Sen, \textit{``{Bound states in gapped graphene
  with impurities: Effective low-energy description of short-range
  interactions}''}, Phys. Rev. \textbf{B78} (2008) 205429,
  [\href{http://arxiv.org/abs/0808.2864}{\texttt{arXiv:0808.2864}}].

\bibitem{nelson2002defects}
David~R. Nelson, \textit{``Defects and geometry in condensed matter physics''};
  Cambridge University Press (2002).

\bibitem{Eguchi:1980jx}
Tohru Eguchi, Peter~B. Gilkey and Andrew~J. Hanson, \textit{``{Gravitation,
  Gauge Theories and Differential Geometry}''}, Phys.\ Rept. \textbf{66} (1980)
  213.

\bibitem{Green:1987sp}
Michael~B. Green, J.H. Schwarz and Edward Witten, \textit{``Superstring theory.
  vol. 1''}; Cambridge Monographs on Mathematical Physics (1988).

\bibitem{Hilbert1933flachen}
David Hilbert, \textit{``\"{U}ber fl\"{a}chen von konstanter gau{\ss}scher
  kr\"{u}mmung''}, in Algebra Invariantentheorie Geometrie, Springer Berlin
  Heidelberg (1933) 437--448.

\bibitem{eisenhart1909treatise}
Luther~Pfahler Eisenhart, \textit{``A treatise on the differential geometry of
  curves and surfaces''}; Ginn and co. (1909).

\bibitem{Chen1993introduction}
C.~Julian Chen, \textit{``Introduction to scanning tunneling microscopy''};
  Oxford University Press, 1~ed. (1993).

\bibitem{Andrei:2012my}
Eva~Y. Andrei, Guohong Li and Xu~Du, \textit{``{Electronic properties of
  graphene: a perspective from scanning tunneling microscopy and
  magneto-transport}''}, Rept. Prog. Phys. \textbf{75} (2012) 056501,
  [\href{http://arxiv.org/abs/1204.4532}{\texttt{arXiv:1204.4532}}].

\bibitem{Maldacena:1997re}
Juan~Martin Maldacena, \textit{``{The large N limit of superconformal field
  theories and supergravity}''}, Int. J. Theor. Phys. \textbf{38} (1999)
  1113--1133.

\bibitem{Witten:1998qj}
Edward Witten, \textit{``{Anti-de Sitter space and holography}''}, Adv. Theor.
  Math. Phys. \textbf{2} (1998) 253--291,
  [\href{http://arxiv.org/abs/hep-th/9802150}{\texttt{hep-th/9802150}}].

\bibitem{Zurek:1996sj}
W.~H. Zurek, \textit{``{Cosmological experiments in condensed matter
  systems}''}, Phys. Rept. \textbf{276} (1996) 177--221,
  [\href{http://arxiv.org/abs/cond-mat/9607135}{\texttt{cond-mat/9607135}}].

\bibitem{Volovik:2000ua}
G.~E. Volovik, \textit{``{Superfluid analogies of cosmological phenomena}''},
  Phys. Rept. \textbf{351} (2001) 195--348,
  [\href{http://arxiv.org/abs/gr-qc/0005091}{\texttt{gr-qc/0005091}}].

\bibitem{Ummarino:2017bvz}
G.~A. Ummarino and A.~Gallerati, \textit{``{Superconductor in a weak static
  gravitational field}''}, Eur. Phys. J. \textbf{C77} (2017), n.~8, 549,
  [\href{http://arxiv.org/abs/1710.01267}{\texttt{arXiv:1710.01267}}].

\bibitem{Ummarino:2019cvw}
G.~A. Ummarino and A.~Gallerati, \textit{``{Exploiting weak field
  gravity-Maxwell symmetry in superconductive fluctuations regime}''}, Symmetry
  \textbf{11} (2019), n.~11, 1341,
  [\href{http://arxiv.org/abs/1910.13897}{\texttt{arXiv:1910.13897}}].

\bibitem{Ummarino2020josephson}
Giovanni~Alberto Ummarino and Antonio Gallerati, \textit{``{Josephson AC effect
  induced by weak gravitational field}''}, Class. Quantum Grav. \textbf{37}
  (2020), n.~21, 217001,
  [\href{http://arxiv.org/abs/2009.04967}{\texttt{arXiv:2009.04967}}].

\bibitem{Gallerati2020interaction}
Antonio Gallerati, \textit{``{Interaction between superconductors and weak
  gravitational field}''}, J. Phys. Conf. Ser. \textbf{1690} (2020) 012141,
  [\href{http://arxiv.org/abs/2101.00418}{\texttt{arXiv:2101.00418}}].

\bibitem{Sepehri:2016nuv}
Alireza Sepehri, Richard Pincak and Ahmed~Farag Ali, \textit{``{Emergence of
  F(R) gravity-analogue due to defects in graphene}''}, Eur. Phys. J. B
  \textbf{89} (2016), n.~11, 250,
  [\href{http://arxiv.org/abs/1606.02039}{\texttt{arXiv:1606.02039}}].

\bibitem{Capozziello:2018mqy}
Salvatore Capozziello, Richard Pincak and Emmanuel~N. Saridakis,
  \textit{``{Constructing superconductors by graphene Chern-Simons
  wormholes}''}, Annals Phys. \textbf{390} (2018) 303--333.

\bibitem{Capozziello:2020ncr}
Salvatore Capozziello, Richard Pin\v{c}ak and Erik Barto\v{s},
  \textit{``{Chern-Simons Current of Left and Right Chiral Superspace in
  Graphene Wormhole}''}, Symmetry \textbf{12} (2020), n.~5, 774.

\bibitem{Baeuerle:1996zz}
C.~Baeuerle, Yu.~M. Bunkov, S.~N. Fisher, H.~Godfrin and G.~R. Pickett,
  \textit{``{Laboratory simulation of cosmic string formation in the early
  Universe using superfluid He-3}''}, Nature \textbf{382} (1996) 332--334.

\bibitem{Ruutu:1995qz}
V.M.H. Ruutu, V.B. Eltsov, A.J. Gill, T.W.B. Kibble, M.~Krusius, Yu.G. Makhlin,
  B.~Placais, G.E. Volovik and Wen Xu, \textit{``{Big bang simulation in
  superfluid He-3-b: Vortex nucleation in neutron irradiated superflow}''},
  Nature \textbf{382} (1996) 334,
  [\href{http://arxiv.org/abs/cond-mat/9512117}{\texttt{cond-mat/9512117}}].

\bibitem{Banados:1992wn}
Maximo Banados, Claudio Teitelboim and Jorge Zanelli, \textit{``{The Black hole
  in three-dimensional space-time}''}, Phys. Rev. Lett. \textbf{69} (1992)
  1849--1851,
  [\href{http://arxiv.org/abs/hep-th/9204099}{\texttt{hep-th/9204099}}].

\bibitem{Cvetic:2012vg}
M.~Cvetic and G.~W. Gibbons, \textit{``{Graphene and the Zermelo Optical Metric
  of the BTZ Black Hole}''}, Annals Phys. \textbf{327} (2012) 2617--2626,
  [\href{http://arxiv.org/abs/1202.2938}{\texttt{arXiv:1202.2938}}].

\bibitem{Kandemir:2019zyt}
B.S. Kandemir, \textit{``{Hairy BTZ black hole and its analogue model in
  graphene}''}, Annals Phys. \textbf{413} (2020) 168064,
  [\href{http://arxiv.org/abs/1907.03509}{\texttt{arXiv:1907.03509}}].

\end{thebibliography}\endgroup

\end{document}